\documentclass[twocol]{ametsocV5}

\usepackage{amsmath,amsfonts,amssymb,bm}
\usepackage{mathptmx}
\usepackage{newtxtext}

\usepackage{multirow}
\usepackage[utf8]{inputenc}
\usepackage{siunitx}
\usepackage{xcolor}
\usepackage[normalem]{ulem}
\usepackage[rightcaption]{sidecap}

\newcommand{\ba}{\begin{eqnarray}}
\newcommand{\ea}{\end{eqnarray}}
\newcommand{\be}{\begin{equation}}
\newcommand{\ee}{\end{equation}}
\newcommand{\benn}{\begin{equation*}}
\newcommand{\eenn}{\end{equation*}}

\title{Robustness of competing climatic states}

\authors{Charline Ragon\thanks{Group of Applied Physics and Institute for Environmental Sciences, University of Geneva, Geneva, Switzerland}, Valerio Lembo\thanks{Institute of Atmospheric Sciences and Climate, Consiglio Nazionale delle Ricerche, ISAC-CNR, Bologna, Italy}, Valerio Lucarini\thanks{Department of Mathematics and Statistics, University of Reading, Reading, UK}\thanks{Centre for the Mathematics of Planet Earth, University of Reading, Reading, UK}, \\ Christian V\'erard\thanks{Section of Earth and Environmental Sciences, University
of Geneva, Geneva,
Switzerland}, J\'er\^ome Kasparian\footnotemark[1] , Maura Brunetti\footnotemark[1]~~\correspondingauthor{M. Brunetti, maura.brunetti@unige.ch}}

\affiliation{}

\abstract{The climate is a non-equilibrium system undergoing the continuous action of forcing and dissipation. Under the effect of a spatially inhomogeneous absorption of solar energy, all the climate components dynamically respond  until an approximate steady state (or attractor) is reached. However, multiple steady states can co-exist for a given forcing and with the same boundary conditions. Here, we apply the Thermodynamic Diagnostic Tool (TheDiaTo) to investigate the statistical properties of  five co-existing climates, ranging from a snowball to an ice-free aquaplanet, obtained in MITgcm coupled simulations. The aim is to explore the multistability of the climate model setup by highlighting differences in competing steady states and their characteristic signatures regarding the meridional transport of heat and water mass, the Lorenz energy cycle and the material entropy production.  
We also investigate how such attractors change when the model configuration is varied. We consider, in particular, the effect of changing the representation of the cloud albedo, and  of implementing an improved closure of the energy budget. We find that, even if the dynamics remains on the same attractor, state variables are modified. The set of metrics in TheDiaTo quantifies such modifications and represents a valuable tool for model evaluation. }

\begin{document}

\maketitle

%
%
%

%








\section{Introduction}

The climate is a highly complex and heterogeneous non-equilibrium system undergoing the continuous action of forcing and dissipation. 
The main source of external forcing is the inhomogenous absorption of incoming solar radiation. The atmosphere dynamically responds to such energy input  by re-distributing the heat from lower to higher values of both latitudes  and altitudes \citep{Peixoto1992,LucariniRevGeo2014,RevModPhys.92.035002}. On the other hand, the ocean is essentially fueled by winds, tides and buoyancy forcing~\citep{MunkWunsch1998,WunschFerrari2004}. As a result, the inhomogeneous absorption of solar heat triggers a complex set of instabilities and feedbacks  occurring at a wide range of temporal and spatial scales, redistributing heat and water mass, 
until approximate steady state conditions are achieved \citep{Peixoto1992} where the dynamics of the system lies upon a high  dimensional attractor \citep{Saltzman2001,LucariniRevGeo2014,RevModPhys.92.035002}. 

A generally accepted theoretical framework is that the climate is a multi-stable dynamical system,  where multiple competing attractors can co-exist under the same forcing and boundary conditions \citep{1969TellA..21..611B,1969JApMe...8..392S,1976JAtS...33....3G,Saltzman2001,RevModPhys.92.035002}. This implies that the phase space is partitioned among the basins of attraction  corresponding to the various attractors and the basin boundaries \citep{2017Nonli..30R..32L}. 
Transition between the competing steady states is possible if the system undergoes forcing of deterministic or stochastic nature~\citep{Saltzman2001,RevModPhys.92.035002,margazoglou2020}.
As an example, observational evidence suggests that during the Neoproterozoic era our planet flipped in and out of a so-called `snowball' state~\citep{Hoffman1998,Pierrehumbert2011}. 
Multi-stability reflects the fact that there are different manners to redistribute the energy among the climate components in such a way that a stable climate can be established. It has been detected in models of different complexity, from energy balance models \citep{1969TellA..21..611B,1969JApMe...8..392S,1976JAtS...33....3G,Abbot2011} and  
general circulation models (gcm) coupled to slab oceans \citep{2010QJRMS.136....2L,2013Icar..226.1724B,Popp2016,2017Nonli..30R..32L,Lucarini2019,Lucarini2020} to fully coupled gcm \citep{2011JCli...24..992F,Rose2015}. All these models have been able to reproduce the dichotomy between the competing warm and snowball climates. But, indeed, modelling exercises indicate the possible existence of additional climatic configurations, such as the slushball Earth \citep{Lewis2007} and the Jormungand state \citep{Abbot2011}.

Recently, up to five attractors were found to co-exist under the same forcing and boundary conditions using the MIT general circulation model (MITgcm) \citep{marshall_finite-volume_1997,marshall_hydrostatic_1997,adcroft_implementation_2004,marshall_atmosphereocean_2004} in a coupled aquaplanet configuration \citep{brunetti2019}. In terms of average surface temperature, these attractors range from the usual snowball state, which is completely covered by ice, to a hot, ice-free state, which is warmer than the usual warm state found in other studies. 
How can this large variety of co-existing states be physically described in a manner that sets the different states apart? To address this issue, we extend the analysis of such attractors performed in \citet{brunetti2019} with the construction of the entire bifurcation diagram for a solar constant ranging between 1336 and 1400~W\,m$^{-2}$ and with the addition of, in particular, two new metrics, the intensity of the Lorenz energy cycle (LEC) and the material entropy production (MEP)  of  the atmosphere.

The extended set of metrics used to characterize the multiple attractors can furthermore be used to quantify biases in climate models,  an important instrument for the understanding of climate dynamics~\citep{Eyring2016,Balaji2017CPMIPMO}. Despite the impressive progress in last decades, biases still remain in the new generation of climate models that are extremely difficult to reduce \citep{Wang2014,Rauser2015,Zhang2015,Palmer2016,Stouffer2017,McKenna2020,Liao2021}. Such biases affect the description of key climatic features like the interplay of global modes of variability~\citep{Yang2018}, their magnitude and frequency, or the occurrence of extremes~\citep{Perkins2011}, and reveal the limits of our ability to correctly reproduce the response to external (solar) or internal (volcanic and anthropogenic) forcings~\citep{Rose2013,Gupta2019}. 

A main source of biases in climate models comes from the use of parameterizations of unresolved processes. In addition to the use of insufficiently observationally constrained parameters, parameterizations
are often not constrained to conservation properties \citep{anthes1985,HourdinEtAl2016}. This can have an impact on the mean properties of the modeled system, inducing numerical drifts, and deserves particular attention. Here, we will focus specifically on two examples, by looking at the effect of different prescriptions for the  cloud albedo and  at the effect of improving the global energy budget of the model, to show the utility of considering a set of metrics including LEC and MEP.

To characterize the different attractors and address model biases in a comprehensive manner we use the Thermodynamic Diagnostic Tool (TheDiaTo, \citet{lembo_2019}).  
TheDiaTo is a software tool that calculates a large set of metrics, including the aforementioned ones, and is thus ideally suited for the task at hand. 
TheDiaTo is also  part of the CMIP6 ESMValTool 2.0 Earth system models diagnostic community effort \citep{gmd-13-3383-2020}. 

The paper is organized as follows. After the description of our setups and simulations used to construct the bifurcation diagram in Section~\ref{sec:2}, we apply TheDiaTo in Section~\ref{sec:3} to study the five competing attractors. 
In Section~\ref{sec:4}, we discuss how re-injecting dissipated kinetic energy affects the physical budgets and the overall properties of the attractors, focusing on the hot state. In Section~\ref{sec:5}, we investigate the impact of changing the cloud albedo scheme on what we refer later to as the cold state, which is intermediate with respect to surface temperature. Finally, we discuss the importance of performing such diagnostics in climate models before drawing our conclusions in Section~\ref{sec:6}.

\section{Methods and simulation setups}
\label{sec:2}

The simulations are performed using the MIT global circulation model (MITgcm, version c65q, \citet{marshall_finite-volume_1997, marshall_hydrostatic_1997, adcroft_implementation_2004}),  a coupled atmosphere-ocean-sea ice model with a 15-levels dynamical ocean, a thermodynamic module for the sea ice component \citep{2000JAtOT..17..525W}, and a 5-layers atmospheric radiative module based on the SPEEDY model \citep{molteni_atmospheric_2003}. The top layer represents the stratosphere, the bottom one the planetary boundary layer, and the remaining three the free troposphere. SPEEDY is an atmospheric module of intermediate complexity that provides a rather realistic representation of the flow despite the coarse vertical resolution and the simplified physical parametrizations, as results from the comparisons with ERA and NCEP/NCAR reanalysis fields~\citep{molteni_atmospheric_2003}, with the advantage of requiring less computer resources than state-of-the-art atmospheric models.

MITgcm is run in an aquaplanet configuration with no continents. The same dynamical kernel is used for both the ocean and the atmosphere, which are represented over the same cubed-sphere grid~\citep{marshall_atmosphereocean_2004}.
Each face of the cube includes $32\times 32$ cells, corresponding to an average horizontal resolution of $2.8^\circ$. The ocean depth is set to a fixed value of 3000~m. The $CO_2$ concentration in the atmosphere is set to 326 ppm and various values for the solar irradiance $S_0$ are considered, as detailed below. The orbital forcing is prescribed at present-day values, thus seasonality is taken into account. Albedos for snow, ice and ocean surface are within the observed range and correspond to those used in \citet{brunetti2019}.
The relative humidity threshold for the formation of low clouds, a parameter denoted as $RHCL2$ in MITgcm, is  set  to the same value for all the simulations considered in the present study, $RHCL2 = 0.7239$.  

\begin{figure}[t]
    \centering
    \includegraphics[width=\columnwidth]{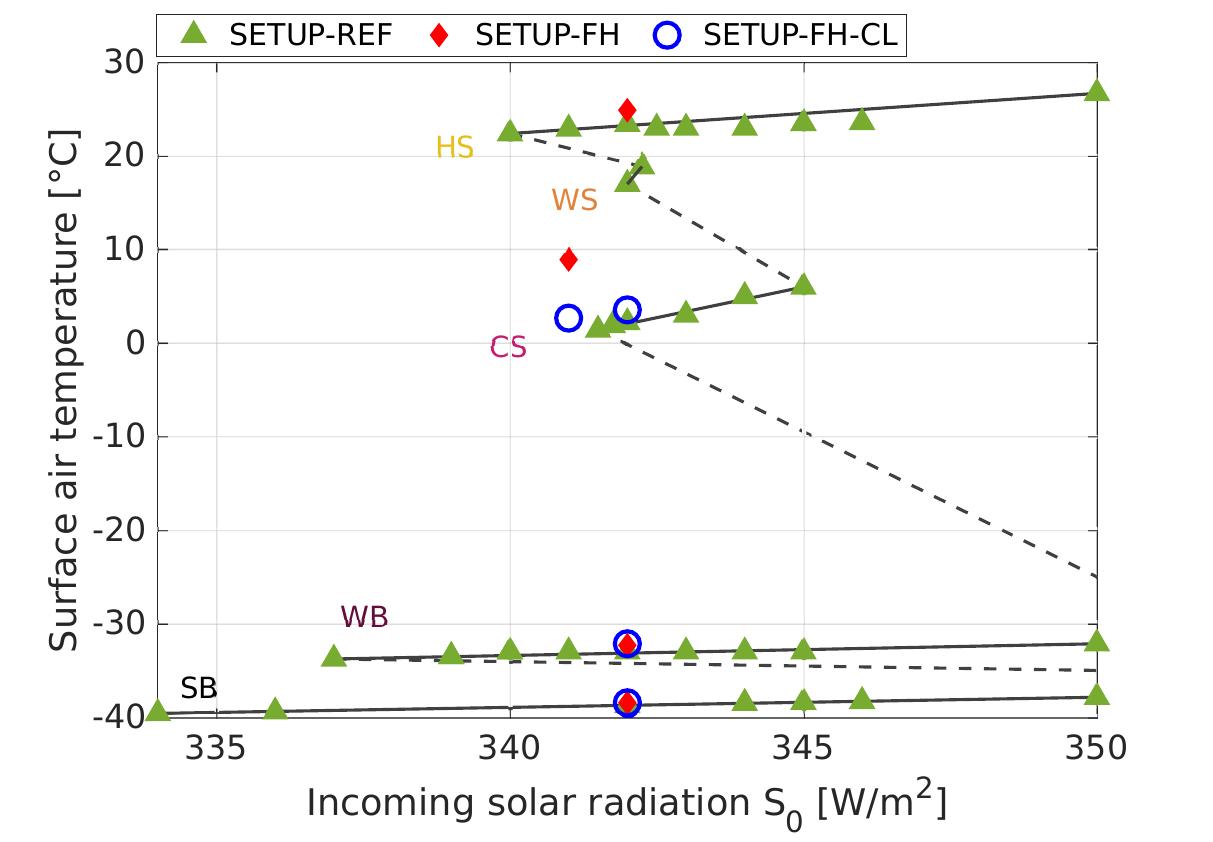}
    \caption{Bifurcation diagram for {\tt SETUP-REF} (green triangles). Solid lines correspond to stable branches, while dashed lines correspond to qualitative sketches of unstable branches. Diamonds and circles denote the attractors for {\tt SETUP-FH} and {\tt SETUP-FH-CL}, respectively.}
    \label{fig:BD}
\end{figure}

We extend the analysis performed in \citet{brunetti2019} by running 28 additional simulations with different incoming solar radiation $S_0$. This allows us to construct the bifurcation diagram\footnote{Other bifurcation diagrams obtained with MITgcm with an horizontal resolution of 3.75$^\circ$ and simplified continental configurations can be found in~\citet{Rose2015,Gupta2019}.} where the mean surface air temperature on the attractors is plotted as a function of the incoming solar radiation in the range $S_0=334-350$~W/m$^2$, as shown in Fig.~\ref{fig:BD} (green triangles). The bifurcation diagram is of critical importance to determine the range of stability of each attractor within a given model setup, and to define multi-stability regions. 
To find the stable branches (solid lines), the simulations are initialised from the five competing attractors found in \citet{brunetti2019} at $S_0 = 342$~W\,m$^{-2}$ with slightly different values of 
incoming solar radiation, and this procedure is repeated until a tipping point is reached, where a shift to a different attractor is observed. 
We take the pragmatic approach of assuming that statistically steady-state conditions are realised within each attractor when its mean annual surface energy balance $F_{\rm{s}}$ becomes lower than $0.2$~\si{W\,m^{-2}} in absolute value, corresponding to an ocean temperature drift 
$dT/dt = F_{\rm{s}}/(c_{\rm{p}}\, \rho\, h)$~\citep[p.~229]{MarshallPlumb2008} lower than $0.05~^\circ$C/Century, with $c_{\rm{p}} = 4000$~J\,K$^{-1}$kg$^{-1}$ the specific heat capacity, $\rho = 1023$~kg\,m$^{-3}$ the sea water density  and $h = 3000$~m the ocean depth. Indeed, under such conditions, we see  virtually no drift in the annual averages of the climatic observables of interest. 
Five stable branches exist, corresponding to the following climates: hot  state (HS),  warm  state (WS),  cold  state (CS),  waterbelt (WB)  and  snowball (SB). In particular, WS is stable within a narrow range of values of the solar constant\footnote{Green triangles are not regularly spaced in Fig.~\ref{fig:BD}, especially near the tipping points $WS\to HS$ and $WS\to CS$ where we have explored in detail the narrow region of stability of WS.}, 1368-1370~W/m$^2$, while SB is stable for all the tested values ranging from 1336 to 1400~W/m$^2$.   
 
The bifurcation diagram is obtained using the reference model setup  where the dissipated kinetic energy is not re-injected in the system and low-cloud albedo is fixed to the constant value  $\alpha_{C0} = 0.38$. This is denoted as {\tt SETUP-REF} (and corresponds to {\tt SetUp2} in \citet{brunetti2019}).  
We then alter one process at a time to better understand the impact of such changes on the model's performance. In {\tt SETUP-FH}, the dissipated kinetic energy is re-injected into the system as thermal energy (see Section~\ref{sec:4}; red diamonds in Fig.~\ref{fig:BD}). This is achieved by means of a flag available in MITgcm, allowing to locally calculate the dissipated kinetic energy at each grid point and vertical level, except the top one representing the stratosphere,  and to re-inject it each time-step as a contribution to the forcing term in the tendency equation for the temperature. In the last considered setup, denoted as {\tt SETUP-FH-CL} (see Section~\ref{sec:5}), we also include a dependence of cloud albedo on latitude~\citep{Kucharski2013} in order to investigate how a different representation of a quantity linked to the transport of both energy and water-mass affects the dynamics on the attractor (open blue circles in Fig.~\ref{fig:BD}).

The energy, water and entropy budgets in the multi-stable states are  
diagnosed with the help of TheDiaTo~\citep{lembo_2019}. After running a simulation for over a thousand years until  steady state conditions are achieved on a given attractor, we continue it for 20 additional years saving  daily and monthly averages of the fields in order to have sufficient statistics for employing the diagnostics. By estimating the energy and water budgets and the corresponding large scale transports, the intensity of the Lorenz Energy Cycle  (LEC) \citep{lorenz_available_1955,Peixoto1992} and the material entropy production (MEP) \citep{Peixoto1992,Goody2000,Pauluis2002,Lucarini2009}, TheDiaTo 
provides a synthetic and physically meaningful characterisation of climate attractors. While details on the calculations of such diagnostics can be found in the original article \citep{lembo_2019}, the main equations are stated for completeness in Appendix~A, together with the interpolation and the steps needed to adapt MITgcm outputs for usage in TheDiaTo.

\section{Characterization of the five aquaplanet climatic attractors}
\label{sec:3}

In order to quantify the relative importance of physical processes in multi-stable states,  this section presents a detailed analysis of the five competing attractors obtained with $S_0=342$~W\,m$^{-2}$ in {\tt SETUP-REF} (as in \cite{brunetti2019}); see 
Fig.~\ref{fig:BD}.   

\subsection{Energy and water-mass budgets and transports} 

Table~\ref{tab:energyBalances_temp} shows a selection of the average values of key global climatic observables for the five competing steady states. Average ocean temperatures range between $T = -1.9~^\circ$C for the SB climate and $T = 17.5~^\circ$C for the HS. Correspondingly, the mean surface air temperature (SAT) ranges between -38.75~$^\circ$C for the  SB climate and 23.2~$^\circ$C for the HS (see Fig.~S1 in Supplemental Material). Let's consider next the temperature difference between polar and equatorial regions (usually referred to as meridional temperature gradient), $\Delta T_{\textrm{PE}}$. This is computed by taking the difference between the average surface air temperature in the latitudinal belt  $[30^\circ~{\rm{S}}, 30^\circ~{\rm{N}}]$ and in the region within $30^\circ$ and $90^\circ$ degrees latitude in the northern and southern hemispheres~\citep{2017Nonli..30R..32L,margazoglou2020}. One finds the smallest value $\Delta T_{\textrm{PE}}$ in the HS (16.2~$^\circ$C), which corresponds to conditions typical of so-called  equitable climates \citep{HUber2011}. Instead, $\Delta T_{\textrm{PE}}$ is largest (34.3~$^\circ$C) for the CS.

\begin{table*}[ht]
\caption{Global mean values averaged over 20 years and associated standard deviation derived from interannual variability for the five attractors at $S_0=342$~W\,m$^{-2}$ in {\tt SETUP-REF}.}
    \label{tab:energyBalances_temp}
    \footnotesize
    \begin{center}
    \begin{tabular*}{\textwidth}{@{\extracolsep\fill}lccccccc}
    \topline
        Description & Name & Units & HS & WS & CS & WB & SB \\
        \midline
         TOA budget & $R_t$ & \si{W\, m^{-2}} & $2.5 \pm ~0.2$ & $2.5 \pm 0.2$ & $2.9 \pm 0.1$ & $1.7 \pm 0.1$ & $0.3 \pm 0.1$ \\
        Surface budget  & $F_{\rm{s}}$ & \si{W\, m^{-2}} &  $0.2 \pm 0.3$ & $-0.0 \pm 0.3$ & $-0.1 \pm 0.1$ & $-0.1 \pm 0.1$ & $(-2\pm~6)\cdot 10^{-5}$  \\
        Ocean drift  & $\partial T/\partial t$ & $^\circ$\si{C/Century} & $0.05 \pm 0.07$ & $-0.01 \pm 0.07$ & $-0.02 \pm 0.02$ & $-0.02 \pm ~0.02$ & $0 \pm 0$ \\
        Water budget & $E - P_\textrm{tot}$ & $10^{-8}$ \si{kg\,m^{-2} s^{-1}} &  $0 \pm 2$ & $0 \pm 1$ & $0 \pm 1$ & $4.3 \pm 0.1$ & $0.00 \pm 0.03$ \\
        Latent heat budget & $R_L$ & \si{W\, m^{-2}} &  $0.01 \pm ~0.05$ & $-0.01 \pm ~0.03$ & $-0.01 \pm ~0.02$ & $0.108 \pm 0.004$ & $-0.0001 \pm 0.0007$ \\
        Total precipitation & $P_\textrm{tot}$ & $10^{-5}$ \si{kg\,m^{-2} s^{-1}} &  $4.47 \pm 0.02$ & $4.08 \pm 0.01$ & $3.190 \pm 0.008$ & $0.435 \pm 0.001$ & $0.1043 \pm 0.0005$ \\
        &&&&&& \\
        Ocean temp.  & $T$ & \si{\celsius} &  $17.515 \pm 0.001$ & $9.990 \pm 0.002$ & $3.223 \pm 0.002$ & $-1.6401 \pm 0.0005$ & $-1.918127 \pm 0.000003$ \\
        Surface air temp.  & SAT & \si{\celsius} &  $23.2 \pm 0.2$ & $17.0 \pm 0.2$ & $2.0 \pm 0.1$ & $-33.05 \pm 0.03$ & $-38.75 \pm 0.04$ \\
         Temp. gradient & $\Delta T_\textrm{PE}$ & \si{\celsius} &  $16.2 \pm 0.2$ & $21.2 \pm 0.2$ & $34.3 \pm 0.2$ & $29.46 \pm 0.08$ & $21.5 \pm 0.1$ \\
        &&&&&& \\
        Mechanical work  & $W$ & \si{W\, m^{-2}} &  $2.43 \pm 0.03$ & $2.42 \pm 0.04$ & $2.06 \pm 0.02$ & $0.72 \pm 0.02$ & $0.37 \pm 0.01$  \\ 
    \botline
    \end{tabular*}
    \end{center}
\end{table*}

\begin{figure}[t]
\centering
    \includegraphics[width=\columnwidth]{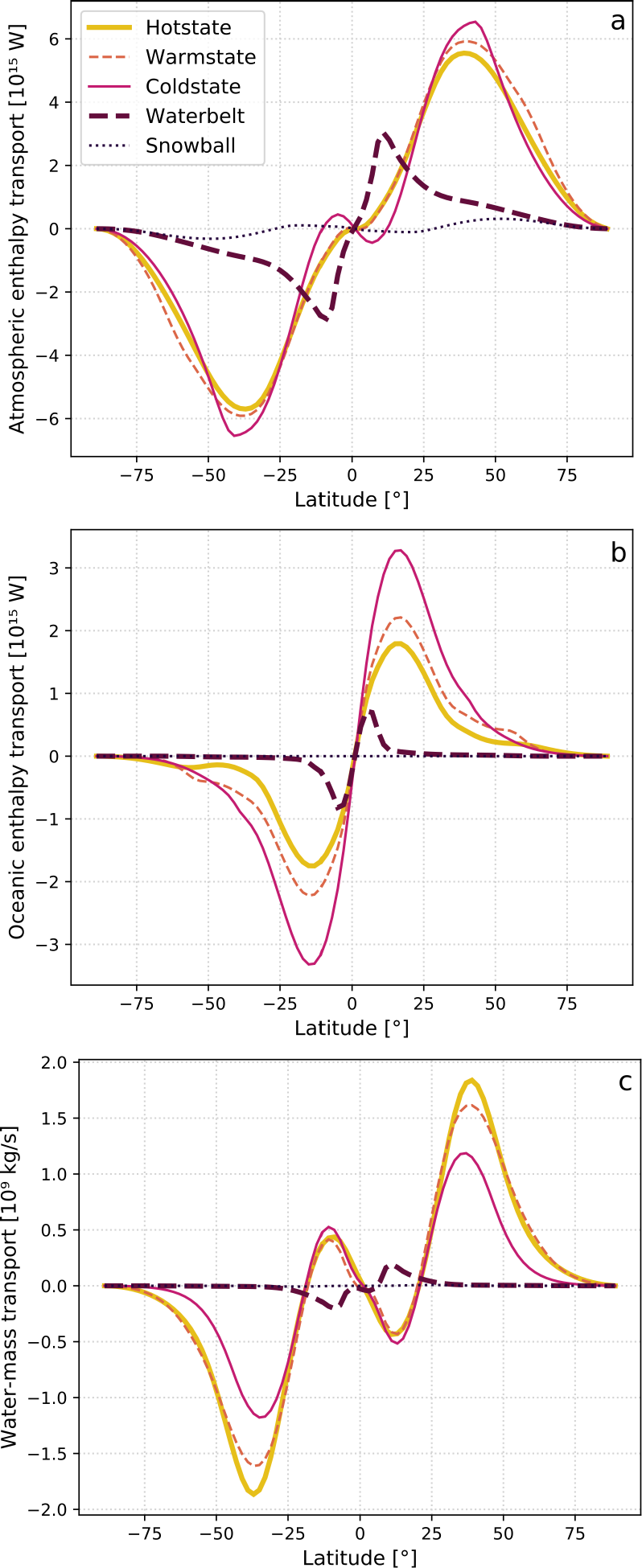}
    \caption{Comparison of northward enthalpy transport in the atmosphere (a) and in the ocean (b), and of northward water-mass transport (c) in the five attractors of {\tt SETUP-REF} for $S_0=342$~W\,m$^{-2}$.}
    \label{fig:enthalpy_transp}
\end{figure}

The Top-Of-Atmosphere (TOA) energy balance ranges between 0.3 and 2.9~W\,m$^{-2}$, which are typical values in coarse resolution simulations such as those performed in the climate models intercomparison project PCMDI CMIP3 (\url{https://esgf-node.llnl.gov/projects/esgf-llnl/}) in the preindustrial scenario (see for example Fig.~2a in \citet{LucariniRagone2011}). 
The presence of an imbalance is in apparent contradiction with steady state conditions; indeed, as discussed in \cite{LucariniRagone2011,Liepert_2012}, the spurious bias is due to physical processes that have been neglected, inconsistently treated or approximated in climate models, as well as to unphysical effects of numerical 
dissipation~\citep{pascale_climate_2011,2012JAMES...4.0A01M,Lauritzen2019,trenberth2020}. In our setting, the main contribution comes from the frictional heating (see Section~\ref{sec:4}) and the fact that sea ice dynamics is neglected~\citep{Brunetti2018}. Note that, as argued in \citet{LucariniRagone2011}, the bias is positive in all cases.

The surface energy budget $F_\textrm{s}$ ranges between $-0.1$ and 0.2~W\,m$^{-2}$ for all climates. These values are in general smaller than those reported for CMIP3 control runs (see Fig.~4 in \citet{LucariniRagone2011}) or for CMIP5 control runs (see Table~2 in \citet{lembo_2019}). The presence of such a small energy imbalance at the  ocean's surface strongly indicates that all our simulations have reached a steady state~\citep{Brunetti2018},  with a drift in the average oceanic temperature of only few parts in $10^{-2}~^\circ$C per century (see Table~\ref{tab:energyBalances_temp}). 
We remark that such accurate steady-state conditions were not imposed, but have been reached after a few  thousand years of simulation, which were required to remove transient effects. 

Table~\ref{tab:energyBalances_temp} also shows the global water budget, given by the difference between global averages of evaporation $E$ and precipitation $P_\textrm{tot}$. We find that in all climates the bias in the water budget is negligible, so that approximately no water is lost during the numerical experiments, in agreement with \citet{Campin2008}. This mirrors the presence of an accurately balanced latent heat budget $R_{\rm{L}}$, also shown in Table~\ref{tab:energyBalances_temp}, which is related to the phase transformation of water over different surfaces; see Appendix~A. For completeness, we also list the global annual mean precipitation $P_\textrm{tot}$, whose maximum value is found in HS.

In all climates, the atmospheric and the oceanic enthalpy\footnote{Following  \citet{LucariniRagone2011,lembo_2019}, the enthalpy transport is given by $\vec J_h = \rho h \vec v$, where the standard definition of enthalpy $h= e+p/\rho$ is adopted, $\rho$ being the density, $e$ the total energy density of the fluid component, $p$ the pressure, and $\vec v$ the velocity vector.} transports (Fig.~\ref{fig:enthalpy_transp}) are poleward and anti-symmetric with respect to the equator as a result of the aquaplanet configuration, which is symmetric between the two hemispheres.  The total transport balances the net TOA radiation influx in the equatorial region and the net TOA outflux in the polar regions \citep{Trenberth2009,LucariniRagone2011}. Going from HS to CS, peaks of meridional enthalpy transport increase in magnitude (Fig.~\ref{fig:enthalpy_transp}), mean surface air temperature decreases and  meridional temperature gradient increases. The enthalpy transport is strongest in CS attractor with a  peak of \SI{6.5}{~\peta\watt} in the atmosphere and \SI{3.3}{~\peta\watt} in the ocean. This is not surprising because the largest part of the atmospheric enthalpy transport in the extratropics is carried out by mid-latitude eddies~\citep[p.~155]{MarshallPlumb2008}, in turn influenced by the meridional temperature gradient, while the oceanic heat transport is proportional to the strength of the circulation multiplied by the temperature gradient at the concerned latitudes~\citep{boccaletti2005}. Since CS has the steepest temperature gradient between pole and equator of the order of $\Delta T_\textrm{PE} = 34.3~^\circ$C, it turns out that it also has the largest meridional enthalpy transport.
Note that, despite the differences between HS, WS and CS, their meridional atmospheric enthalpy transport peaks at the same latitude ($\approx 40^\circ$~N/S) in agreement with the classical prediction by \cite{Stone1978}. The oceanic enthalpy transport is in all cases less intense than the atmospheric one and peaks at a lower latitude, as in the present-day Earth climate. 

The much colder WB and SB climates are fundamentally different from the previous three in terms of transport profiles. In the WB climate, the atmospheric enthalpy transport peaks at the boundary of the water belt, where a large meridional temperature gradient is locally realised. Instead, the peak of the oceanic transport is obtained even closer to the equator and results from the intense overturning circulation inside the water belt. The meridional transports are negligible in the ice covered portion of the planet. In the case of the SB climate, the oceanic transport vanishes due to the absence of an ice-free ocean surface, and the atmospheric transport is extremely weak at all latitudes, 
which, in turn, agree with the fact that a) the meridional temperature differences are very small and that b) the atmosphere is dry, see Fig.~\ref{fig:enthalpy_transp}c.

In the case of the three warmer climates, the annual mean meridional moisture transport - see Fig.~\ref{fig:enthalpy_transp}c - is qualitatively similar to the present one, with the peak of the poleward transport occurring where the intensity of the meridional eddies is strongest - namely, in the storm track corresponding to the peak of the meridional enthalpy transport - whereas an equatorward transport is realised in the tropical region, coincident with the Hadley cells. Note that the largest poleward transport is obtained in the HS climate because higher surface temperature favours evaporation. Considering that the HS has the weakest total enthalpy transport (Fig.~\ref{fig:enthalpy_transp}a),  one deduces that large scale latent heat transport is relatively more important in the HS. Clearly, meridional moisture transport is almost vanishing in the SB climate, because evaporative processes are virtually absent, and is very weak and concentrated over and near the water belt in the WB climate.

\subsection{Lorenz Energy Cycle (LEC)}

The Lorenz energy cycle (LEC) describes the time-averaged transformation of energy between the available potential form and the kinetic form. The reservoir of available potential energy $P$ is continuously replenished thanks to the inhomogeneous absorption of radiation, while the kinetic energy $K$ is continuously depleted as a result of dissipative processes \citep{lorenz_available_1955}. This is the starting point for treating the atmosphere as a non-ideal engine, for defining its efficiency, and evaluating its entropy production \citep{Peixoto1992,Goody2000,Pauluis2002,Lucarini2009}. 

\begin{figure}[t]
    \centering
    \includegraphics[width=\columnwidth]{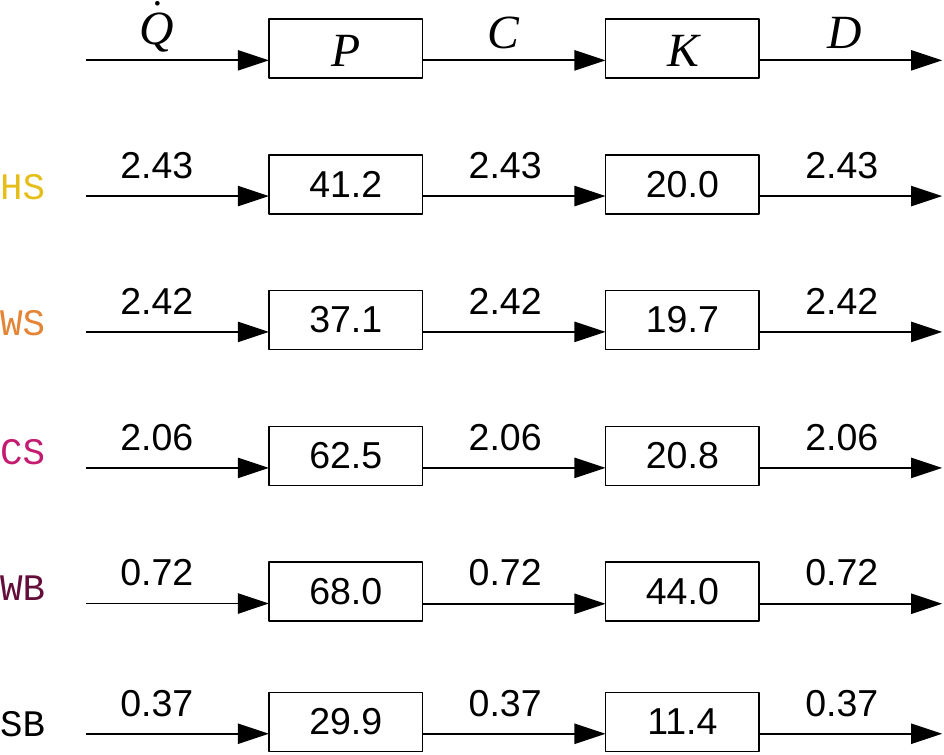}
    \caption{Simplified Lorenz energy cycle for the five attractors in {\tt SETUP-REF} for $S_0=342$~W\,m$^{-2}$. Storage terms (boxes) of available potential energy $P$ and kinetic energy $K$ are in $10^5$~J\,m$^{-2}$; generation $\dot Q$, conversion $C$ and dissipation $D$ terms (arrows) are in W\,m$^{-2}$. Standard deviations, derived from interannual variability and associated to each value from HS to SB, are: (0.03, 0.04, 0.02, 0.02, 0.01)~W\,m$^{-2}$ for $\dot Q$, $C$ and $D$; (0.8, 0.5, 0.9, 0.5, 0.4)$\cdot 10^5$~J\,m$^{-2}$ for $P$ and (0.2, 0.2, 0.1, 0.4, 0.5)$\cdot 10^5$~J\,m$^{-2}$ for $K$.}
    \label{fig:LECsimplified}
\end{figure}

Specifically, a simplified version\footnote{The more general version of LEC is able to distinguish between processes occurring at different scales of motion and to describe energy exchanges across scales \citep{lorenz_available_1955,Peixoto1992}; see Supplemental Material.} of the LEC can be represented in terms of two energy reservoirs and three conversion terms as in~\cite{pascale_climate_2011}, where the available potential energy reservoir $P$ evolves due to generation by diabatic heating $\dot Q$ and conversion of potential to kinetic energy $C$. On the other hand, the kinetic energy reservoir $K$ is affected by the conversion term $C$ and the dissipation term $D$. For sake of clarity, the LEC can be formally summarised by the following budget equations:
\ba
    \dot P &=& -C(P,K) + \dot Q
    \label{eq:pascale_LEC1}\\
    \dot K &=& -D + C(P,K) 
    \label{eq:pascale_LEC2}
\ea
The definitions of reservoir and conversion terms are given in Supplemental Material and in \cite{lembo_2019}.

\begin{figure*}[t]
\includegraphics[width=\textwidth]{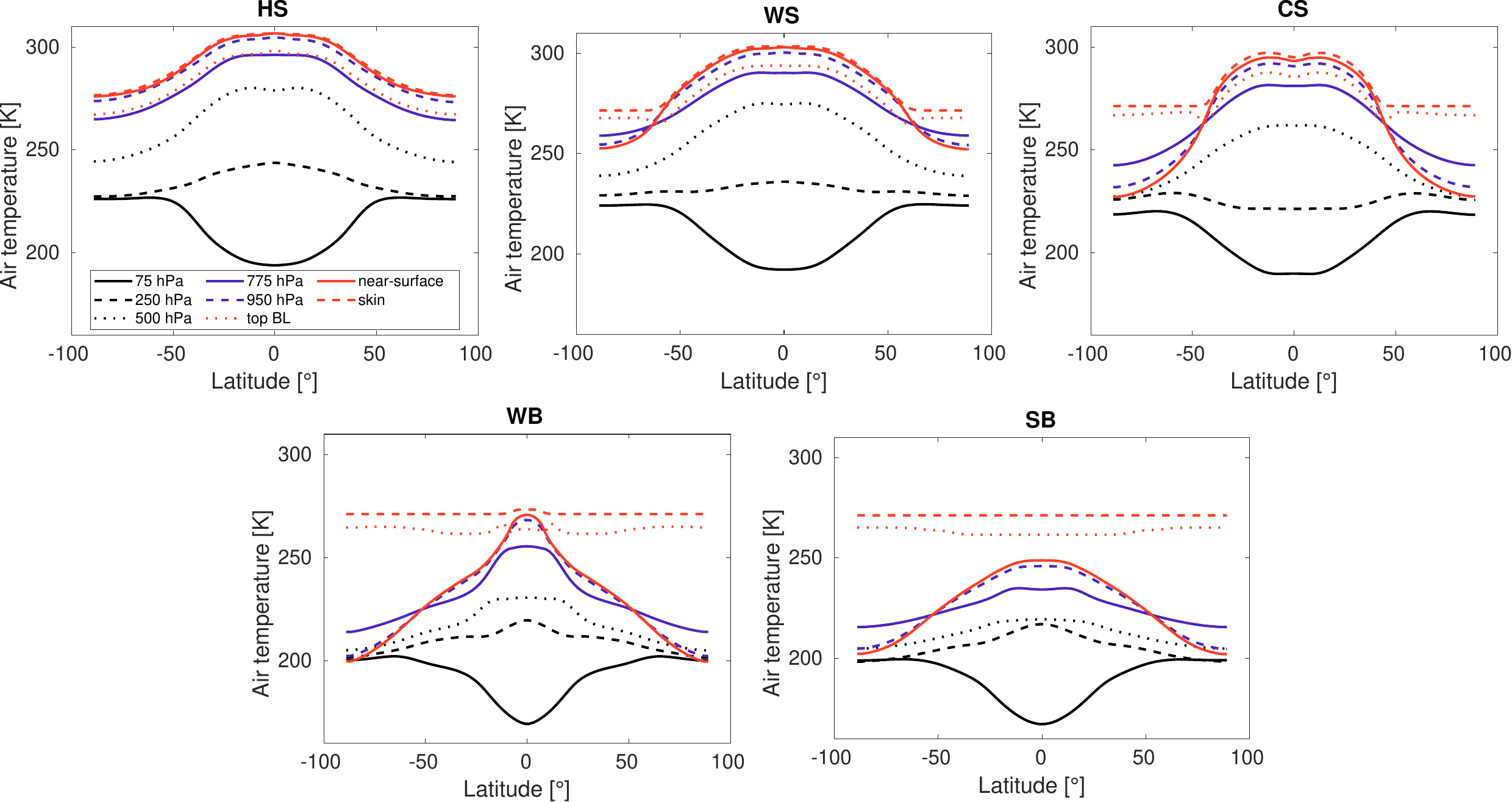}
\caption{Zonal average of air temperature at different pressure levels in the five attractors in {\tt SETUP-REF} for $S_0=342$~W\,m$^{-2}$.}
\label{fig:airtemp}
\end{figure*} 

\begin{figure*}[t]
\includegraphics[width=\textwidth]{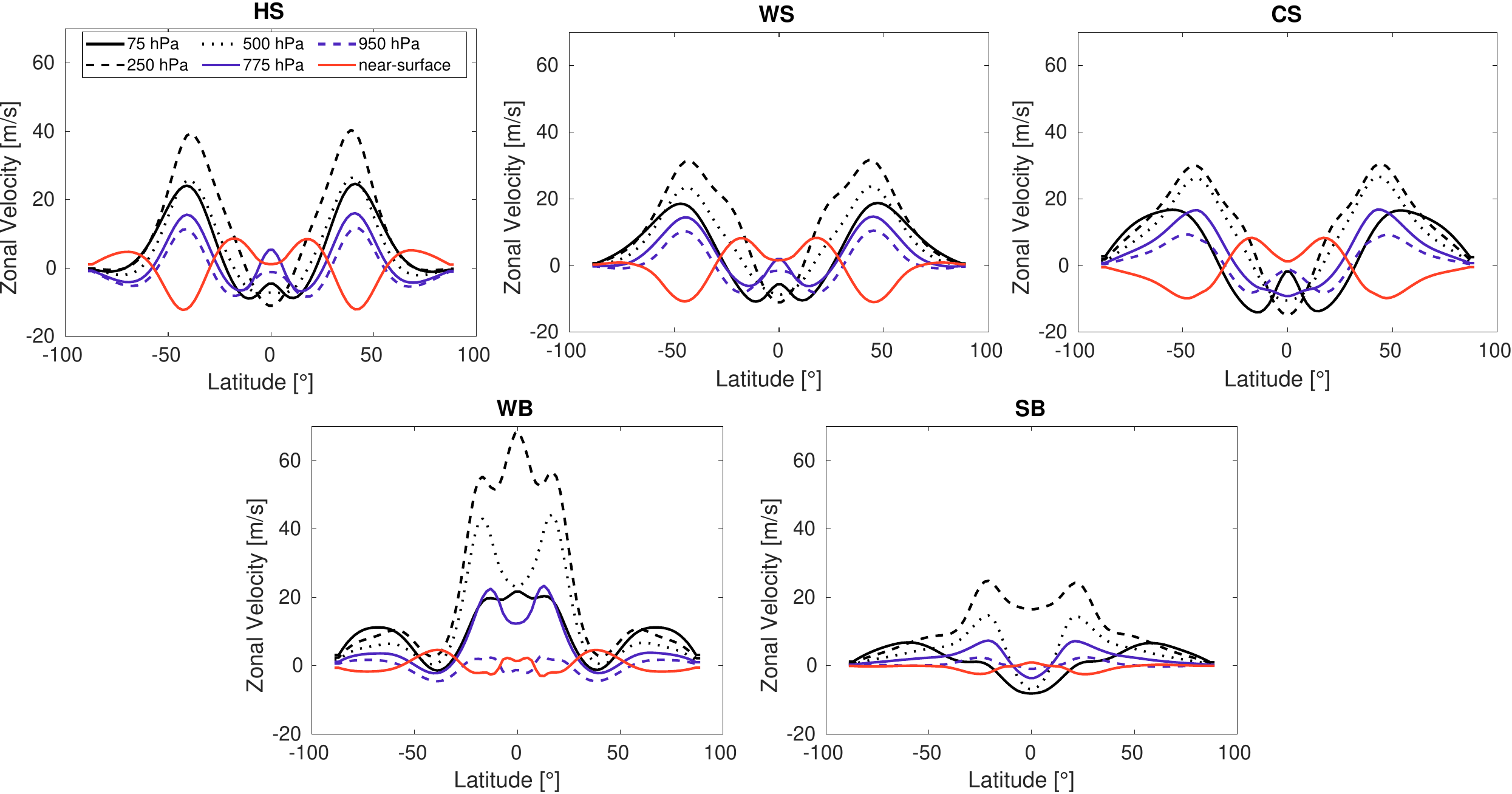}
\caption{Zonal wind at different pressure levels in the five attractors in {\tt SETUP-REF} for $S_0=342$~W\,m$^{-2}$.}
\label{fig:zonalwind}
\end{figure*} 

The reservoir of available potential energy $P$ is smaller in 
HS and WS than in CS, in agreement with the fact that, as far as simulations of the actual Earth are concerned, it  decreases from pre-industrial to present-day conditions~\citep{lembo_2019}. $P$ is proportional to the variance of temperature on isobaric surfaces in the free troposphere, and to the inverse of the difference in the lapse rate of dry {\it vs.} moist air, $\Gamma_d-\Gamma$~\citep[see eq.~(10) in that paper]{lorenz_available_1955}. The former is dominated by the variance with respect to the zonal mean (see Supplemental Material), in turn related to meridional temperature gradients in the free atmosphere (Fig.~\ref{fig:airtemp}), while $\Gamma_d-\Gamma$ is larger in warmer conditions~\citep[p.~49]{MarshallPlumb2008}. The combined effect of such contributions can explain the values of $P$ in the attractors. The fact, for example, that $P$ in HS is larger than in WS depends on weaker meridional gradients in WS at 250, 500 and 775~hPa, as shown in Fig.~4.  On the other hand,  the reservoirs of kinetic energy $K$, depending on the variance of wind speed, are very similar for HS, WS and CS (Fig.~\ref{fig:zonalwind}). Consequently, the sum of kinetic and available potential energy is different between the attractors. 

The WB climate features the largest reservoirs of both available potential and kinetic energy. Indeed, the presence of a very intense temperature gradient localized at the ice edge (Fig.~\ref{fig:airtemp}) and governing the dynamics~\citep{brunetti2019} leads to large values of the available potential energy and allows for the presence of very intense zonal winds  at low latitudes and at intermediate pressure levels, as shown in Fig.~\ref{fig:zonalwind}.  In this attractor, jet streams are so intense that the mean kinetic energy reservoir amounts to almost 0.65 times the mean potential energy reservoir. The SB climate, instead, features the smallest reservoirs for both forms of energy as a result of the weak temperature gradients and weak atmospheric circulations throughout the domain. 
 
Statistically steady state conditions imply that the long-term time-average values of the diabatic heating $\dot Q$, of the conversion term $C$ (which can also be seen as the mechanical work $W$ performed by the climatic engine or as the intensity of the LEC), and of the dissipation $D$ are identical. This is indeed the case for the values obtained  for the five competing climatic attractors (Fig.~\ref{fig:LECsimplified} and Table~\ref{tab:energyBalances_temp}). The intensity of the LEC is found to be considerably higher for the three warmer climates, where it ranges between 2.06 and 2.43~W\,m$^{-2}$. These figures are in broad agreement with the values obtained with seven climate models participating in CMIP5 using pre-industrial conditions \citep[Table~2]{lembo_2019}, as well as with those found in additional CMIP5 model runs and in reanalysis datasets \citep{Veiga2013,Li2007}.
In contrast to the warmer climates, the intensity of the LEC is much weaker in the WB and in the SB climates, where weather variability is greatly reduced and localised in a narrow band (WB) or virtually absent (SB); see \cite{2010QJRMS.136....2L} for a detailed analysis of the thermodynamics of the SB state. Note that, as well known~{\citep{lorenz_available_1955}},  there is no obvious relationship between the size of the energy reservoirs and the value of the conversion terms, so the fact that the WB state has the largest  reservoirs of energy is not in contradiction with the very low intensity of the LEC. 

\subsection{Material-entropy production (MEP)}

\begin{table*}[ht]
\begin{center}
\caption{Contributions to material-entropy production in the five attractors at  $S_0=342$~W\,m$^{-2}$ in {\tt SETUP-REF}.}
    \label{tab:MEP}
    \small
    \begin{tabular*}{\textwidth}{@{\extracolsep\fill}lccccc}
    \topline
        MEP [\si{mW\,m^{-2}K^{-1}}] associated to... & Hot state & Warm state & Cold state & Waterbelt & Snowball \\
        \midline
        Viscous processes & $8.3\pm 0.1$ & $8.4\pm 0.1$ & $7.32\pm 0.08$ & $2.70\pm 0.06$ & $1.39\pm 0.04$ \\
        Hydrological cycle & $47\pm 4$ & $42\pm 2$ & $30\pm 1$ & $2.7\pm 0.2$ & $0.89\pm 0.09$ \\
        {\hspace{0.3cm} \it Evaporation} & $-371\pm 2$ & $-342\pm 1$ & $-272.0 \pm 0.6$ & $-40.3\pm 0.1$ & $-9.62\pm 0.04$ \\
        {\hspace{0.3cm} \it Potential energy of droplets} & $6.30\pm 0.05$ & $5.50\pm 0.03$ & $3.77\pm 0.01$ & $0.386\pm 0.001$ & $0.1194\pm 0.0004$ \\
        {\hspace{0.3cm} \it Precipitation} & $412\pm 2$ & $379\pm 1$ & $298.7\pm 0.7$ & $42.6\pm 0.1$ & $10.39\pm 0.05$ \\
        Sensible heat diffusion & $1.07\pm 0.03$ & $1.55\pm 0.02$ & $2.409\pm 0.007$ & $2.558\pm 0.007$ & $2.547\pm 0.006$ \\
        {\bf Total MEP} & $\mathbf{56\pm 4}$ & $\mathbf{52\pm {2}}$ & $\mathbf{40\pm 1}$ & $\mathbf{8.0\pm 0.3}$ & $\mathbf{4.8\pm 0.1}$ \\
    \botline    
    \end{tabular*}
    \end{center}
\end{table*}

In the Earth system energy undergoes transport and transformation processes, and entropy is continuously produced by irreversible mechanisms. At steady state, while a zero net energy flux is reached at the boundary of the system, the entropy production is balanced by a net outgoing flux of entropy towards space. The entropy production can be partitioned into two main components~\citep{LucariniRevGeo2014,Bannon2015}. One - the dominant part - describes the irreversible transformation of the radiation from short-wave to infrared. 
The second part, usually referred to as material entropy production, is 
generated by fluid motions mainly through: 1)
dissipation of kinetic energy due to viscous processes; 2) irreversible processes associated with moisture; 3) sensible heat fluxes at the interface between the atmosphere and the surface~\citep{Peixoto1992,Goody2000,Pauluis2002,Pauluis2007,10.1175/2011JAS3713.1}.
In the current climate, term 3) is the smallest, followed by term 1), whereas irreversible moist processes give by far the most relevant contribution to the total MEP \citep{Goody2000,pascale_climate_2011,LP2014}.  Moist processes are embedded in the hydrological cycle, namely the phase changes - {\it e.~g.} evaporation, condensation, and sublimation taking place in non-saturated environment - and the dissipation of kinetic energy from precipitating hydrometeors (see eq.~(\ref{eq_app:MEP})).

For sake of simplicity, TheDiaTo neglects the phase
changes occurring within the clouds during the formation and depletion of rain/snow droplets. Moreover, it deliberately focuses on MEP related to irreversible processes in the atmosphere, as it was previously shown~\citep{pascale_climate_2011} that the contribution of the ocean to the MEP budget is at least one order of magnitude smaller than the atmospheric contribution.

Table~\ref{tab:MEP} lists individual contributions to MEP for the five attractors. The total MEP increases with the mean surface  air temperature ({\it i.~e.}, from SB to HS) \citep{2010QJRMS.136....2L} mainly because the hydrological cycle is stronger in a warmer environment~\citep{HeldSoden2006} - see also  Fig.~\ref{fig:enthalpy_transp}c. 
In contrast, the contribution coming from the diffusion of sensible heat is larger in colder climates, albeit it only represents a small fraction of the total MEP.
This quantity depends on the product of the sensible heat flux, $H_\textrm{S}$, and the difference between top-of-boundary layer and skin temperatures, $T_\textrm{BL} -T_\textrm{s}$ (see second term in eq.~(\ref{eq_app:MEP})). As can be seen in Fig.~\ref{fig:airtemp}, it turns out that, while $T_\textrm{BL} -T_\textrm{s}$ is of the same order in all attractors, the sensible heat flux $H_\textrm{S}$ is larger in colder climates, since it depends on the difference between skin and near-surface temperatures~\citep[p.~226]{MarshallPlumb2008}. 

\subsection{Summary}

\begin{figure}[t]
\includegraphics[width=\columnwidth]{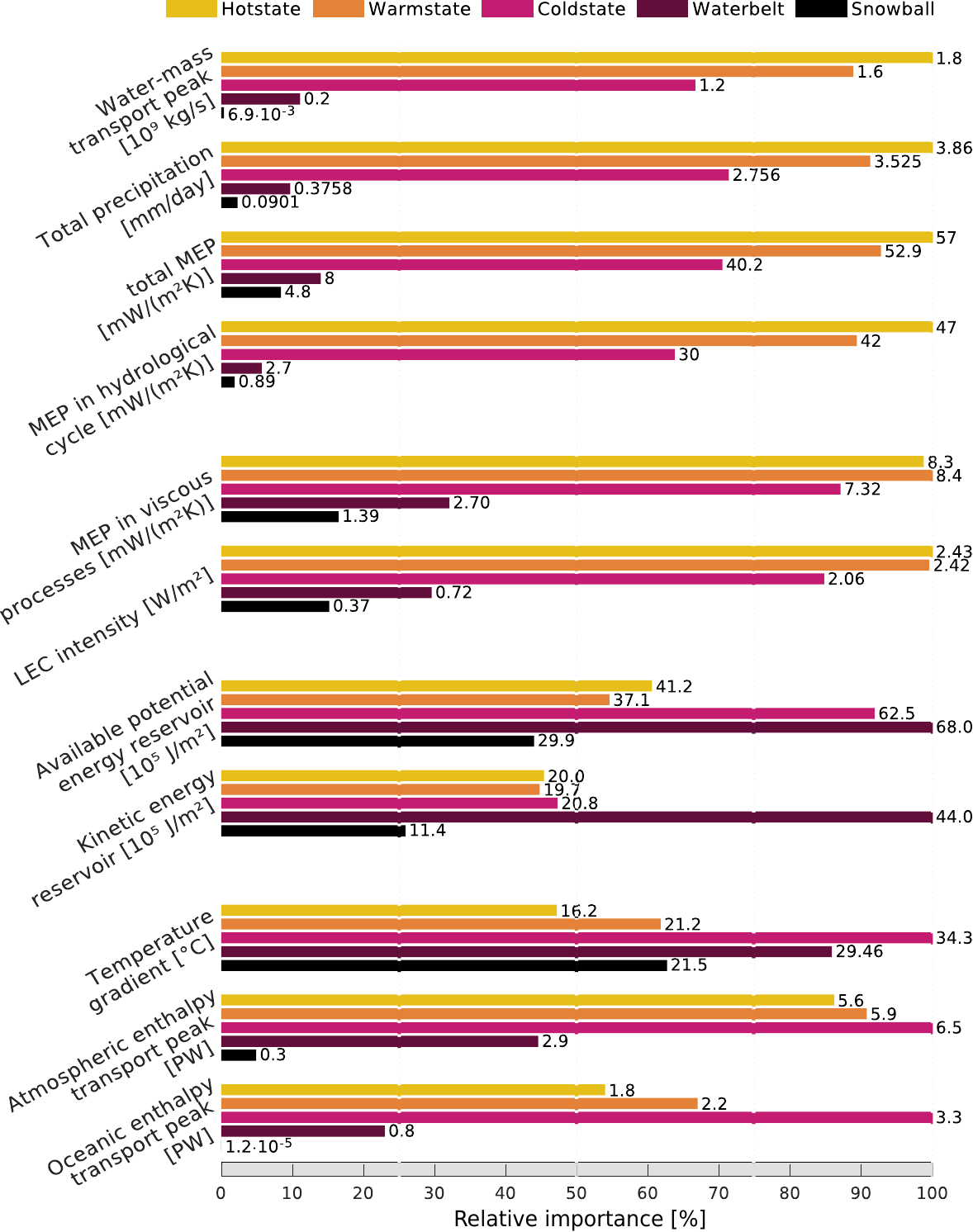}
\caption{Summary of the main climatic variables and  of their values in the five  states in {\tt SETUP-REF} for $S_0=342$~W\,m$^{-2}$. The relative value is expressed with respect to the largest one realized among the five states.}
\label{fig:summary}
\end{figure} 

The analysis performed so far is  summarised in Fig.~\ref{fig:summary}, where for each considered climatic variable the relative value is expressed with respect to the largest one realized among the five states. This visually highlights how, under the same forcing, the processes balance in a different way in co-existing attractors. It also helps in understanding how different variables are connected between each other,  and which are the key quantities that identify a climatic state. In particular, we see that MEP due to the hydrological cycle (that represents the largest contribution to total MEP) is linked to total precipitation and the peak intensity in the water-mass transport. These quantities, together with the global temperature, are maximised in the HS, an ice-free steady state.  On the other hand,  
the LEC intensity is closely linked to MEP due to viscous processes, which is largest in the WS. The meridional temperature gradient influences the enthalpy transport in both the ocean and the atmosphere, showing the most prominent peaks in the CS. Localised gradients of temperature and wind speed in the free troposphere increase the variance of  such quantities, thus maximising values of $P$ and $K$, respectively, as in WB. Finally, SB minimises all the above quantities. The inclusion of LEC and MEP is crucial to distinguish the behaviour of WB and WS from the other steady states, thus showing the relevance of using TheDiaTo for attractors characterisation. 

\section{Effect of removing the energy bias by the re-injection of the dissipated kinetic energy}
\label{sec:4}

\begin{table*}[ht]
\caption{Comparison of hot states at 342~\si{W\,m^{-2}} in {\tt SETUP-REF} and {\tt SETUP-FH}. Statistically different values are in bold.}
    \label{tab:STD_FH}
    \begin{center}
    \begin{tabular}{ccccc} 
    \topline
        Description & Name & Units & {\tt SETUP-REF}  & {\tt SETUP-FH}\\
        \midline
        Surface air temperature & $SAT$ & \si{\celsius} & $\mathbf{23.2 \pm 0.2}$ & $\mathbf{24.92 \pm 0.08}$ \\
        Meridional temperature gradient & $\Delta T_\textrm{PE}$ & \si{\celsius} & $\mathbf{16.2 \pm 0.2}$ & $\mathbf{15.44 \pm 0.09}$ \\
        TOA energy budget & $R_\textrm{t}$ & \si{W\,m^{-2}} & $\mathbf{2.5 \pm 0.2}$ & $\mathbf{0.0 \pm 0.2}$ \\
        Surface energy budget & $F_\textrm{s}$ & \si{W\,m^{-2}} & $0.2 \pm 0.3$ & $0.2 \pm 0.2$ \\
        Water budget & $E-P_\textrm{tot}$ & $10^{-8}$~\si{kg\,m^{-2} s^{-1}} & $0 \pm 2$ & $0 \pm 2$ \\
        Total precipitation & $P_\textrm{tot}$ & $10^{-5}$~\si{kg\,m^{-2} s^{-1}} & $\mathbf{4.47 \pm 0.02}$ & $\mathbf{4.56 \pm 0.02}$ \\
        Mechanical work & $W$ & \si{W\,m^{-2}} & $\mathbf{2.43 \pm 0.03}$ & $\mathbf{2.33 \pm 0.03}$ \\
    \botline   
    \end{tabular}
    \end{center}
\end{table*}

\begin{table*}[ht]
\caption{Contributions to material entropy production in hot states with {\tt SETUP-REF} and {\tt SETUP-FH} at 342~W\,m$^{-2}$. Statistically different values are in bold.}
    \label{tab:FH_MEP}
    \begin{center}
    \begin{tabular}{lcc}
    \topline
        MEP [\si{mW\,m^{-2}K^{-1}}] associated to... & {\tt SETUP-REF} & {\tt SETUP-FH}\\
        \midline
        Viscous processes & $\mathbf{8.3\pm 0.1}$ & $\mathbf{7.9\pm 0.1}$ \\ 
        Hydrological cycle & $47\pm 4$ & $48\pm 3$\\ 
        \hspace{0.3cm} {\it Evaporation} & $\mathbf{-371\pm 2}$ & $\mathbf{-377\pm 1}$ \\
        \hspace{0.3cm} {\it Potential energy of droplets} & $\mathbf{6.30\pm 0.05}$ & $\mathbf{6.53\pm 0.03}$ \\
        \hspace{0.3cm} {\it Precipitation} & $\mathbf{412\pm 2}$ & $\mathbf{419\pm 2}$ \\
        Sensible heat diffusion & $\mathbf{1.07\pm 0.03}$ & $\mathbf{0.91\pm 0.02}$ \\
        {\bf Total MEP} & $56\pm 4$ & $57\pm 3$ \\
    \botline
    \end{tabular}
    \end{center}
\end{table*}

\begin{figure}[ht!]
    \centering
    \includegraphics[width=\columnwidth]{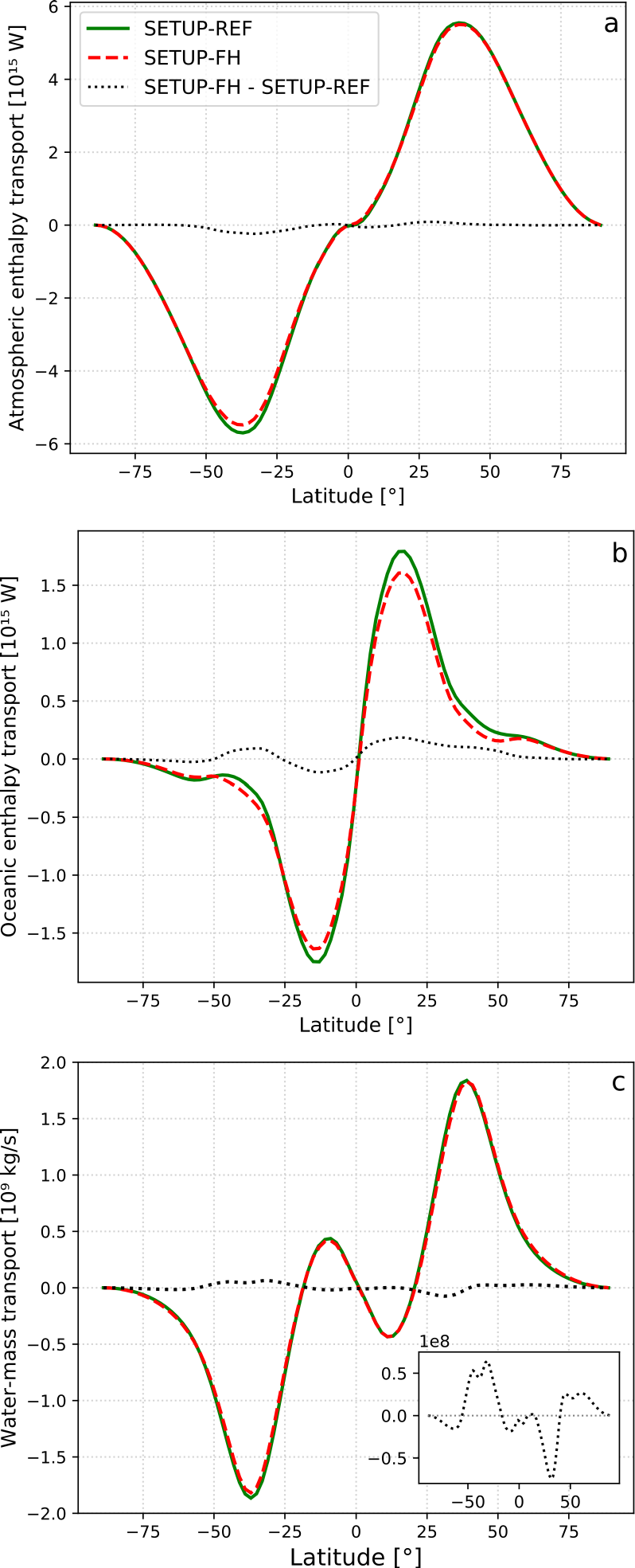}
    \caption{Comparison of hot states at $S_0 = 342$~W\,m$^{-2}$ in {\tt SETUP-REF} and {\tt SETUP-FH} in terms of northward enthalpy transport in the atmosphere (a) and in the ocean (b), and water-mass transport (c), where the inset is a zoom of the difference.}
    \label{fig:FHtransports}
\end{figure}

\begin{figure}[ht!]
    \centering
    \includegraphics[width=\columnwidth]{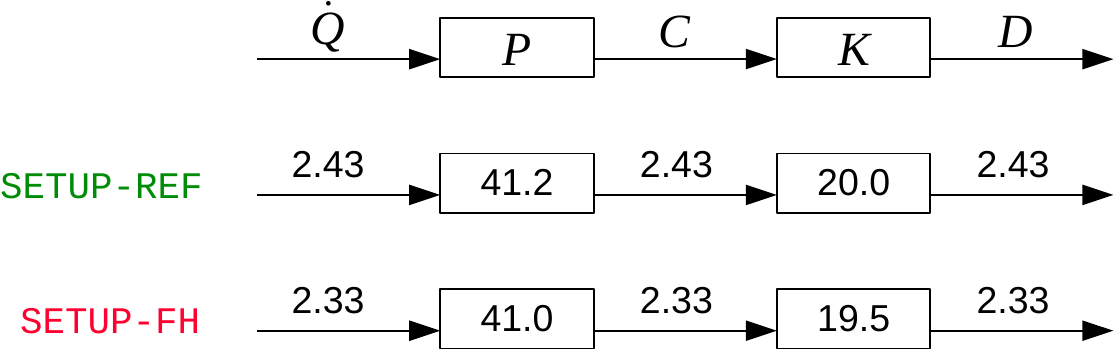}
    \caption{Simplified Lorenz energy cycle for hot states at $S_0 = 342$~W\,m$^{-2}$ in {\tt SETUP-REF} and {\tt SETUP-FH}. Storage terms (boxes) in $10^5$~J\,m$^{-2}$ and conversion terms (arrows) in W\,m$^{-2}$. Standard deviations are: $0.03$~W\,m$^{-2}$ for $\dot Q$, $C$ and $D$; $0.8\cdot 10^5$~J\,m$^{-2}$ for $P$; $0.2\cdot 10^5$~J\,m$^{-2}$ for $K$, in both setups. }
    \label{fig:fh_lec}
\end{figure}

After having characterized the attractors, we can now evaluate their robustness against different model configurations.
Kinetic energy within eddies is dissipated at small scales through an energy cascade ({\it i.~e.}, the dissipation term denoted as $DE$ in the complete LEC; see Supplemental Material). Energy conservation imposes that such dissipated mechanical energy should enter again into the energy budget as it is eventually converted into internal energy by friction. This term is usually ignored in general circulation models, as it is much smaller than other contributions, such as the latent heat exchanges. However, one should keep in mind that the frictional dissipation is positive definite, and, hence, does play a role in the overall energy budget. It has been shown that neglecting this term gives rise to a spurious thermal forcing of up to 2~W\,m$^{-2}$~\citep{becker_frictional_2003} and could explain part of the bias observed in TOA energy imbalance for the climate models~\citep{LucariniRagone2011,wild_global_2020}. Thus, we have performed a sensitivity experiment, in which we evaluate the impact of the re-injection of dissipated kinetic energy on the modeled energy budget and the other thermodynamic diagnostics computed in TheDiaTo.    
In order to assess the relevance of this effect, we focus on the HS,  comparing the standard setup {\tt SETUP-REF} and the energy-consistent setup {\tt SETUP-FH} for $S_0 = 342$~W\,m$^{-2}$; see Figure~\ref{fig:BD}.

While the imbalances of water mass and surface energy are similar in both cases, as shown in Table~\ref{tab:STD_FH}, the TOA imbalance is almost exactly reduced to zero (within the confidence interval) when frictional heating is re-injected as in {\tt SETUP-FH}, confirming the importance of including such term in climate simulations. This improved conservation of energy gives rise to an increase of mean surface air temperature of 1.72~$^\circ$C  (see Table~\ref{tab:STD_FH}) and an average increase of 1.4~$^\circ$C in the five atmospheric layers, as conjectured in~\citet{LucariniRagone2011}.

The transport of enthalpy is slightly less intense in {\tt SETUP-FH}, as shown in Fig.~\ref{fig:FHtransports}a, b. The energy re-injection is local in space and time and takes place mostly in the mid-latitudes, which is where the strongest dissipation occurs. Hence, heat is added in a region where the annual-mean TOA budget is negative (see Fig.~S3 in Supplemental Material). As a result, the peak value of the meridional enthalpy transport as well as the meridional temperature gradient $\Delta T_\textrm{PE}$ decrease in {\tt SETUP-FH}, as shown in Fig.~\ref{fig:FHtransports}a, b and Table~\ref{tab:STD_FH}, respectively. Instead, the local increase in the surface temperature leads, as a result of enhancement of evaporation, to a slight strengthening of the meridional moisture transport in {\tt SETUP-FH} with respect to the reference setup {\tt SETUP-REF} (Fig.~\ref{fig:FHtransports}c). 

The LEC for the two setups are compared in Fig.~\ref{fig:fh_lec}. 
Dissipation $D$ is 4\% smaller in {\tt SETUP-FH} than in the reference setup.  
In a steady state, the generation of available potential energy $\dot Q$  and the dissipation $D$ have to balance and hence $\dot Q$ has to decrease by an equal amount. 
This is consistent with a weaker meridional enthalpy transport in the atmosphere, as observed in Fig.~\ref{fig:FHtransports}a, and with a smaller meridional temperature gradient (from equator to poles), see Table~\ref{tab:STD_FH}.

While the intensity of the atmospheric circulation is lower in {\tt SETUP-FH}, as measured by the mechanical work $W\sim D$, its hydrological cycle becomes slightly more effective, as already observed for the water-mass transport in Figure~\ref{fig:FHtransports}c, with statistically significant larger values of material-entropy production associated with precipitation and condensation, as shown in Table~\ref{tab:FH_MEP}. Heating  the lower levels of the atmosphere strengthens the hydrological cycle and favours vertical transport of water vapour.  Overall, the total MEP remains approximately constant  because the slightly strengthening of the hydrological cycle in {\tt SETUP-FH} is compensated by the reduced contribution of viscous processes and sensible heat diffusion.

In summary, re-injecting frictional heating improves the TOA budget, as expected, but has also several additional consequences, that have been revealed by TheDiaTo: on one hand, the increased mean temperature of the atmospheric column, particularly over the mid-latitudes, has strengthened  the total precipitation and MEP associated to it; on the other hand, the reduced dissipation of energy has weakened the mechanical work of the LEC and the diabatic heating related to the meridional temperature gradient, and consequently the meridional enthalpy transport. 

\section{Testing different cloud albedo representations} 
\label{sec:5}

\begin{table*}[ht]
  \caption{Comparison of cold states at 341~\si{W\,m^{-2}} in {\tt SETUP-FH} and {\tt SETUP-FH-CL}. Statistically different values are in bold.}
    \label{tab:FH_FHCLOUDS}
    \begin{center}
    \begin{tabular}{ccccc}
    \topline
        Description & Name & Units & {\tt SETUP-FH}  & {\tt SETUP-FH-CL}  \\
        \midline
        Surface air temperature & $SAT$ & \si{\celsius} & $\mathbf{8.93 \pm 0.08}$ & $\mathbf{2.67 \pm 0.09}$ \\
        Meridional temperature gradient & $\Delta T_{\textrm{PE}}$ & \si{\celsius} & $\mathbf{27.3 \pm 0.2}$ & $\mathbf{33.5 \pm 0.2}$ \\
        Sea ice extent & & $10^6$~\si{km^2} & $\mathbf{118.7 \pm 0.8}$ & $\mathbf{159 \pm 1}$ \\
        TOA energy budget & $R_\textrm{t}$ & \si{W\,m^{-2}} & $-0.3 \pm 0.2$ & $-0.3 \pm 0.2$ \\
        Surface energy budget & $F_\textrm{s}$ & \si{W\,m^{-2}} & $0.0 \pm 0.2$ & $0.0 \pm 0.2$ \\
        Water budget & $E-P_\textrm{tot}$ & $10^{-9}$~\si{kg\,m^{-2}s^{-1}} & $-5 \pm 8$ & $-5 \pm 8$ \\
        Total precipitation & $P_\textrm{tot}$ & $10^{-5}$~\si{kg\,m^{-2}s^{-1}} & $\mathbf{3.537 \pm 0.009}$ & $\mathbf{3.168 \pm 0.009}$ \\
        Mechanical work & $W$ & \si{W\,m^{-2}} & $1.98 \pm 0.03$ & $2.02 \pm 0.02$ \\
        \botline
    \end{tabular}
    \end{center}
\end{table*}

\begin{table*}[ht]
\caption{Contributions to material entropy production in cold states with {\tt SETUP-FH} and {\tt SETUP-FH-CL} at 341~\si{W/m^2}. Statistically different values are in bold.}
    \label{tab:CLOUDS_MEP}
    \begin{center}
    \begin{tabular}{lcc}
    \topline
        MEP [\si{mW\,m^{-2}K^{-1}}] associated to... & {\tt SETUP-FH} & {\tt SETUP-FH-CL}\\
        \midline
        Viscous processes & $7.0\pm 0.1$ & $7.18\pm 0.09$ \\
        Hydrological cycle & $\mathbf{35\pm 2}$ & $\mathbf{30 \pm 2}$ \\
        \hspace{0.3cm} {\it Evaporation} & $\mathbf{-299.4\pm 0.7}$ & $\mathbf{-270.0\pm 0.7}$ \\
        \hspace{0.3cm} {\it Potential energy of droplets} & $\mathbf{4.39\pm 0.02}$ & $\mathbf{3.75\pm 0.01}$ \\
        \hspace{0.3cm} {\it Precipitation} & $\mathbf{330.1\pm 0.8}$ & $\mathbf{296.6\pm 0.8}$ \\
        Sensible heat diffusion & $\mathbf{2.04\pm 0.02}$ & $\mathbf{2.275\pm 0.008}$ \\
        {\bf Total MEP} & $\mathbf{44\pm 2}$ & $\mathbf{39\pm 2}$ \\
        \botline
    \end{tabular}
    \end{center}
\end{table*}

\begin{figure}[ht!]
    \centering
    \includegraphics[width=\columnwidth]{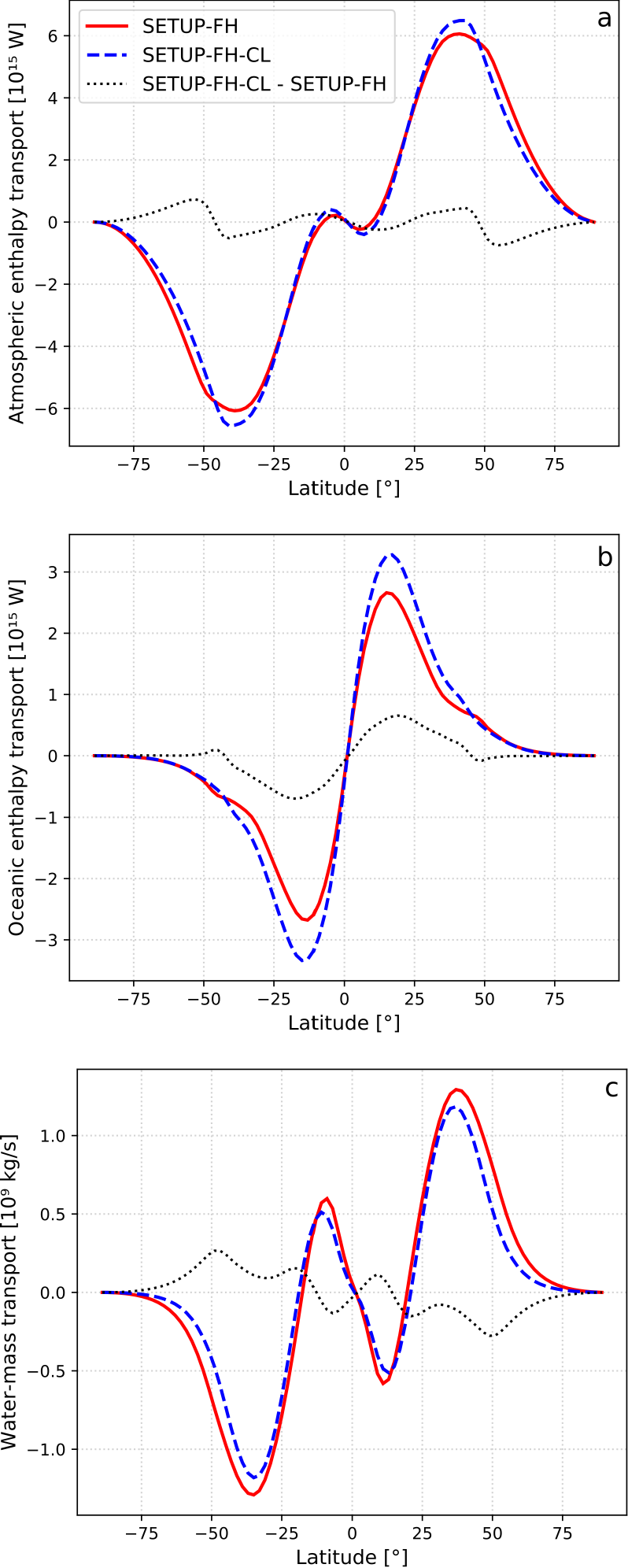}
    \caption{Comparison of cold states at $S_0=341$~W\,m$^{-2}$ in {\tt SETUP-FH} and {\tt SETUP-FH-CL} in terms of enthalpy transport in the atmosphere (a) and in the ocean (b), and of water-mass transport (c).}
    \label{fig:CLOUDtransports}
\end{figure}

\begin{figure}[ht!]
    \centering
    \includegraphics[width=\columnwidth]{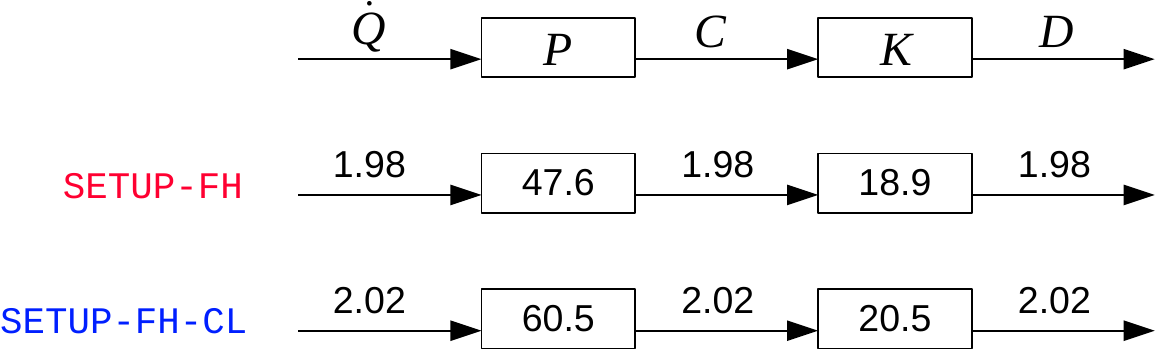}
    \caption{Simplified Lorenz energy cycle for cold states at 341~\si{W\,m^{-2}} in {\tt SETUP-FH} and {\tt SETUP-FH-CL}. Storage terms (boxes) in $10^5$ \si{J\,m^{-2}} and conversion terms (arrows) in \si{W\,m^{-2}}. Standard deviations are: $0.03$/$0.02$~W\,m$^{-2}$ for $\dot Q$, $C$ and $D$; $0.8$/$0.8 \cdot 10^5$~J\,m$^{-2}$ for $P$; $0.2$/$0.1\cdot 10^5$~J\,m$^{-2}$ for $K$, in {\tt SETUP-FH/SETUP-FH-CL}, respectively. }
    \label{fig:clouds_lec}
\end{figure}

A new representation for the cloud albedo has been introduced in the latest version of the atmospheric module SPEEDY (called ICTPAGCM, \citet{kucharski_decadal_2006,Kucharski2013}) in order to reduce the net solar radiation at high latitudes and, hence, to have better agreement  with observational data.   
The cloud albedo is assumed to depend on the latitude $\phi$:
\begin{equation}
\alpha_C(\phi) = \begin{cases} \alpha_{C0} + 0.2*|\sin(\phi)|^4  & \textrm{if}~\cos(\phi) > \frac{1}{2} \\ 
\alpha_{C0} + 0.2*|\sin(\arccos(1/2))|^4 & \textrm{elsewhere}, \end{cases}   
\end{equation}
so that $\alpha_C(\phi)$ increases poleward from a value of $\alpha_{C0} = 0.38$ at the equator up to a maximum value of $\approx0.4925$ at 60$^\circ$S/60$^\circ$N, and then holds constant up to the poles. 

The physical rationale behind such a representation is that cloud cover appears effectively thicker when radiation is coming from lower geometrical angles. We have thus included such a modification in the atmospheric module of MITgcm and we call this configuration {\tt SETUP-FH-CL}.

If we consider an incoming solar radiation of 342~~\si{W/m^2}, the WB and SB states are observed in both {\tt SETUP-FH} and {\tt SETUP-FH-CL} (see Fig.~\ref{fig:BD}). We will not investigate the effect of changing the cloud albedo representation for these two climates because they feature very weak dynamical processes and because they are in deep frozen state with high-albedo surface at high-latitudes, so that the effect of changing the albedo of high-latitude clouds can be understood as minimal.

In order to compare the effects of the cloud albedo prescription on attractors with stronger mechanical work, we consider  an incoming solar radiation of 341~\si{W/m^2} and look at the  cold climates realized with {\tt SETUP-FH} and {\tt SETUP-FH-CL}; see red diamond and blue circle in Fig.~\ref{fig:BD}\footnote{Note that for this value of the incoming solar radiation the HS becomes unstable in {\tt SETUP-FH} and morphs into a climate that is close to the CS in {\tt SETUP-FH-CL}.}.

Table~\ref{tab:FH_FHCLOUDS} gives the key output characterizing the attractor in both  configurations. First of all, it is important to verify that the new representation affects neither the global water-mass budget nor the energy budget. The increase of cloud albedo at high latitudes reduces the amount of incoming radiation, and the mean surface air temperature by nearly \SI{6}{\celsius}. This is correlated to a larger sea ice extent in {\tt SETUP-FH-CL}. 
However, both values of the sea ice cover are of the same order (120-160 millions of km$^2$) as those found in the cold climates at $S_0 = 342$~W/m$^2$ (see Table~2 in \citet{brunetti2019}), suggesting that both attractors are of the same type.
As a result of the more uneven absorption of solar radiation between high and low latitudes, the transports of enthalpy in both the atmosphere and the ocean (Figs.~\ref{fig:CLOUDtransports}a, b) are stronger in {\tt SETUP-FH-CL}. Moreover, the larger sea-ice extent determines an equatorward displacement of the peaks in the atmospheric transport. Consistently with what was found in Section~\ref{sec:3}, the {\tt SETUP-FH-CL} climate has a higher reservoir of available potential energy $P$ (see Fig.~\ref{fig:clouds_lec}) and a weaker moisture transport with its main peaks located closer to Equator,
as shown in Fig.~\ref{fig:CLOUDtransports}c. Indeed, the average temperature of the planet is greatly reduced with the new high-latitude cloud albedo prescription, hence the atmosphere as a whole becomes much drier, and a larger sea-ice extent reduces the ocean surface available for water evaporation. 

The total MEP turns out to be significantly lower in {\tt SETUP-FH-CL}, as a result of a smaller contribution in all components of the MEP budget, except for MEP associated to sensible heat diffusion and viscous processes, as shown in Table~\ref{tab:CLOUDS_MEP}. This is due to the weaker  hydrological cycle in {\tt SETUP-FH-CL}, confirming the trend observed in Section~\ref{sec:3} for attractors with colder temperatures. 
The change in MEP associated with viscous processes is not statistically significant, and the same holds for the intensity of LEC that is comparable in the two setups, of the same order as in CS at $S_0 = 342$~W/m$^2$  in {\tt SETUP-REF} or {\tt SETUP-FH-CL}, and lower than in WS (see Table~\ref{tab:FH_FHCLOUDS} and Fig.~\ref{fig:clouds_lec}), 
showing that the attractor is robust with respect to the new cloud prescription.

All in all, when looking at the CS, including the new representation leads to a decrease in the  total MEP by $\approx 11.4$\% due to a weaker hydrological cycle, and strengthens the enthalpy transport in both the atmosphere and the ocean, without introducing any spurious bias in water and energy budgets.
Since \citet{kucharski_decadal_2006}  showed that such cloud albedo prescription improves comparisons with observational data in present-day simulations, we can conclude that it is worth including it in the MITgcm atmospheric module.

\section{Summary and conclusions}
\label{sec:6}

The climate is a nonequilibrium, multiscale system that features multistability. The occurrence of such phenomenon arises from its complex dynamics where forcing, dissipative processes, and nonlinear feedbacks can balance each other in different ways. The presence of multistability is intimately connected to the existence of tipping points coming with qualitative changes in the system dynamics for suitably defined forcings. In a deterministic setting, multistability is described by the presence of more than one competing steady state associated with different attractors, each included in a separate basin of attraction. While the dynamics of an autonomous system is confined to a single attractor, as its evolution is uniquely defined by its initial condition, the presence of stochastic forcings makes it possible for the system to explore  the full phase space by performing transitions between the various basins of attraction \citep{Benzi1983, Saltzman2001,Lucarini2019,Lucarini2020,RevModPhys.92.035002}.

We performed simulations using the MIT general circulation model with coupled aquaplanet configuration and different values of the incoming solar radiation $S_0$ in order to construct the bifurcation diagram (Fig.~\ref{fig:BD}). Such methodology and the resulting graph are crucial to identify multistable regions, tipping points and the range of stability of each attractor. 
The model displays five attractors that co-exist for solar constant values in the range 1368-1370~W\,m$^{-2}$ (see Fig.~\ref{fig:BD}): the classical "snowball" (SB) and "warm state" (WS) solutions, the very cold waterbelt (WB) state, where a water belt is present near the equator, the "cold state" (CS), where sea-ice extends to the mid-latitudes, and the "hot state" (HS), an equitable climate where no sea ice is present at all. 

In order to describe the properties of such competing climatic states, we made use of a newly developed diagnostics tool, TheDiaTo~\citep{lembo_2019},  that is part of the most recent version of the ESMValTool suite for evaluating Earth system models \citep{gmd-13-3383-2020}. TheDiaTo provides flexible tools for the characterisation of fundamental thermodynamic and conservation laws, starting from first principles. More specifically, we have focused on global averages and patterns of near-surface temperatures, water-mass, TOA and surface energy budgets, hence describing the strength of the meridional enthalpy and moisture transports, as well as the atmospheric Lorenz energy cycle (LEC) through its reservoirs, sources, sinks and conversion terms. The second law of thermodynamics has been assessed through the retrieval of material entropy production (MEP), {\it i.~e.} the entropy change related to irreversible processes, such as the energy exchanges through sensible heat fluxes and the hydrological cycle. We have found the key processes that characterise each attractor: total precipitation, peak intensity in water-mass transport and global temperature are maximised in HS; LEC intensity and MEP in viscous processes are large in WS; meridional temperature gradient and enthalpy transport are dominant in CS; $P$ and $K$ are maximised when localised gradients of temperature and wind speed are present, as in WB; SB minimises all the previous quantities. These findings provide hints for using different climatic variables on reduced phase space for each climatic state, a subject that we will explore in further studies. 

Exploring the multi-stability of the climate system requires to consider a wide range of initial conditions and detailed characterisation of the resulting climates. Such framework is of great relevance for modelling paleoclimate conditions~\citep{2014CliPa..10.2053P,BrunettiVerard2015,Ferreira2018GRL,MessoriFaranda2020}, where large uncertainty exists in initial conditions, as well as for the study of exoplanets, where the probability of finding an habitable planet is increased when multistability is allowed~\citep{Seager2013,2013Icar..226.1724B}. Characterizing the basic properties of the modelled dynamics and thermodynamics is thus crucial, in order to suitably represent the tipping points of the climate system \citep{Lenton2008} and allow for a more general definition of climate sensitivity~\citep{vonDerHeydt2017,Ashwin2020}.

The complete set of metrics in TheDiaTo can be used not only to quantitatively differentiate the attractors but also to highlight biases due to a simplified representation of physical processes in climate models. For testing this last point, we performed two numerical experiments.
First, focusing on the HS, we forced re-injection of frictionally dissipated kinetic energy into the overall energy budget. On one hand, such correction effectively leads to a reduction in the TOA energy bias, as expected. On the other hand, it influences the dynamics of the atmosphere in a non-trivial way, as the re-injection is co-located with the frictional dissipation. The climate becomes warmer, the meridional temperature gradient is reduced, as well as the LEC intensity, while the total precipitation and MEP associated to it increase. Second,  focusing on the CS, we investigated the effect of introducing a dependency on latitude in the cloud albedo (cf.~\citet{kucharski_decadal_2006,Kucharski2013}).  
The implementation of  such a  scheme does not affect global conservation properties  but causes changes in energy and moisture fluxes that lead to a cooler climate, with a weaker hydrological cycle and MEP.  Despite these changes, TheDiaTo allowed us to verify that the attractors are of the same type (CS), showing that useful information can be inferred on the attractors nature even without the construction of the complete bifurcation diagram.  

In summary, in both the numerical experiments the set of metrics in TheDiaTo has clearly shown that the climate state slightly changes, remaining on the same attractor, with non-negligible impacts on the dynamics and thermodynamics and, at the same time, preserving or improving the global conservation properties. Thus, TheDiaTo helps in controlling simulation quality and in evaluating different configurations. In particular, LEC and MEP,  two metrics that are not routinely used for model evaluation, can provide additional benchmarks when constrained by observational data in present-day climate simulations.

While TOA energy imbalance can be improved though the inclusion of missing physical processes (such as frictional heating) or enhanced  algorithms (with less numerical  dissipation, for example), the energy imbalance at the surface $F_s$ is less affected by such procedures. As it is shown in several previous diagnostic studies, the \textit{ghost} energy bias is concentrated in the atmosphere \citep{LucariniRagone2011,LucariniRevGeo2014,Lembo2017}. Instead, $F_s$ stabilizes as one considers longer simulation time, which allows the ocean to reach an approximate steady state. A vanishing surface energy imbalance (as  assured in our simulations) guarantees that the drift of the mean ocean temperature  is negligible. Thus, TOA and surface energy imbalance should be always monitored in climate models~\citep{Brunetti2018} and we encourage to explicitly list them when presenting simulation results.

At present, TheDiaTo allows to perform a thermodynamic analysis of the atmosphere only. However, in principle, it is possible to establish a Lorenz energy cycle also for energy exchanges and transformations within the ocean~\citep{Storch2013}. As high resolution coupled models now allow for resolutions that are consistent with explicitly resolved mesoscale ocean eddies, a successive development would be creating a set of diagnostics that include the dynamics and thermodynamics of the oceans specifically, and it is left for future work.


\bigskip
{\it Acknowledgements.} 

We are grateful to Nicolas Roguet, Anar Artan, Antoine Vandendriessche, Antoine Branca and Mathieu Fanetti for running some of the MITgcm simulations. We thank anonymous reviewers for very useful remarks that helped in improving the manuscript. The computations were performed on the Baobab cluster at University of Geneva. C.~R., C.~V., J.~K. and M.~B. acknowledge the financial support from the
Swiss National Science Foundation (Sinergia Project CRSII5\_180253).
V.~Lembo was supported by the Collaborative Research Centre
TRR181 `Energy Transfers in Atmosphere and Ocean' funded by the Deutsche Forschungsgemeinschaft
(DFG, German Research Foundation), project No. 274762653. V. Lucarini acknowleges the support provided by the Horizon 2020 project TiPES (grant no. 820970).  

%
%
\datastatement
The datasets generated during the current study are available from the corresponding author upon  request.

%






%
%
%

\appendix[A]
\appendixtitle{Main equations used in TheDiaTo}

 The goal of this Appendix is to remind the reader how to calculate the main quantities discussed in this paper and to describe approximations/adaptations used to apply TheDiaTo on MITgcm outputs. We use (as much as possible) the same notation as~\citet{lembo_2019}, where the reader can find all the details for the derivation of the equations reported here.

\subsection{Energy budget and transport}

Radiative fluxes at the surface ($F_s$) and at TOA ($R_t$) depend on latitude $\phi$, longitude $\lambda$ and time $t$ as follows:
\begin{eqnarray}
    \label{eq_app:budg_surf}
    F_s(\phi, \lambda, t) &=& S_\textrm{s}^\downarrow - S_\textrm{s}^\uparrow + L_\textrm{s}^\downarrow - L_\textrm{s}^\uparrow - H_\textrm{S}^\uparrow - H_\textrm{L}^\uparrow \\
    R_\textrm{t}(\phi, \lambda, t) &=& S_\textrm{t}^\downarrow - S_\textrm{t}^\uparrow - L_\textrm{t}^\uparrow
    \label{eq_app:budg_toa}
\end{eqnarray}
where $S$ is shortwave radiation, $L$ longwave radiation, $H_\textrm{L}$ is latent heat flux and $H_\textrm{S}$ is the sensible heat flux. Subscripts $t$ and $s$ denote top-of-atmosphere and surface, respectively. Upward ($\uparrow$) or downward ($\downarrow$) direction is also shown. 

Global energy imbalances are computed by averaging $F_\textrm{s}$ and $R_\textrm{t}$ over the total surface and over a period of time of 20 years.

The meridional transport $\mathcal{T}$ is computed by taking the long-term temporal and zonal averages of eqs.~(\ref{eq_app:budg_surf})-(~\ref{eq_app:budg_toa}), as follows:
\begin{equation}
    \mathcal{T}(\phi) = 2 \pi \int_{\phi}^{\pi/2} a^2 \cos \phi' <\overline{F(\phi',\lambda, t)}>  \,d\phi'
    \label{eq_app:transport}
\end{equation}
where $F$ is the radiative flux (at surface, $F_\textrm{s}$, or at TOA, $R_\textrm{t}$), 
$a$ is the Earth's radius, $< >$ represents the long-term time mean and overline the zonal mean. The atmospheric transport is computed as the difference between the transport at TOA and that at surface, $\mathcal{T}_\textrm{atm} = \mathcal{T}_\textrm{TOA} - \mathcal{T}_\textrm{s}$.

\subsection{Budget and transport of water-mass and latent heat}

The water-mass budget in the atmosphere corresponds to the difference between global averages of surface evaporation and precipitation, $E - P_{\textrm{tot}}$, where $P_{\textrm{tot}}$ includes both convective and large-scale precipitation.
Since rainfall and snowfall precipitations are not differentiated in MITgcm, these are both accounted within $P_{\textrm{tot}} = P_\textrm{r} + P_\textrm{s}$. 

The evaporation $E$ is an output of MITgcm and its calculation  is based on different evaporation coefficients for ocean, land (when it is present) and ice surfaces. Thus, evaporation is not calculated from latent heat as proposed in TheDiaTo (see eq.~(6) in that paper). 

The meridional water-mass transport is obtained by calculating a cumulative integral over latitude of the zonal mean and long-term time mean of $E-P_{\textrm{tot}}$, analogously to eq.~(\ref{eq_app:transport}).

The latent heat budget $R_\textrm{L}$ depends on the latent heat flux $H_\textrm{L}^\uparrow$ and precipitation $P_{\textrm{tot}}$:
\begin{equation}
    R_\textrm{L} = H_\textrm{L} - L_\textrm{v} P_{\textrm{tot}} 
\end{equation}
where $L_\textrm{v} = 2.5008 \cdot 10^6 ~\si{J/kg}$ is the latent heat of evaporation.
In MITgcm, the latent heat flux is deduced as a residual between the total energy surface budget (an output of MITgcm called {\tt TFLUX} = $F_\textrm{s}$) and all the other components (as in eq.~(\ref{eq_app:budg_surf})). Note that in this way the heat associated with snow melting is taken into account in $H_\textrm{L}$, assuring in general a well closed budget also for the latent heat, as shown in Table~\ref{tab:energyBalances_temp}.

\subsection{Material entropy production}

In order to compute the material entropy production, we use the `direct method' of TheDiaTo, where all contributions from irreversible processes are explicitly 
estimated as follows: 
\begin{eqnarray}
    && \textrm{MEP} = 
    \underbrace {\int_A \frac{\kappa_\textrm{s}}{T_\textrm{d}} dA }_{\textrm{Viscous processes}}
    - \underbrace { \int_A H_\textrm{S} \left (\frac{1}{T_\textrm{s}} - \frac{1}{T_{\textrm{BL}}} \right) dA }_{\textrm{Sensible heat diffusion}}\nonumber \\
    &&- \underbrace { \int_A \frac{L_\textrm{v} ~E}{T_\textrm{s}} dA}_{\textrm{Evaporation}}
    + \int_{A_\textrm{r}} \left( \underbrace { \frac{L_{\textrm{v}}~P_{\textrm{tot}}}{T_{\textrm{c}}}}_{\textrm{Precipitation}} + \underbrace { g \frac{P_{\textrm{tot}}~h_{\textrm{ct}}}{T_\textrm{p}}}_{\textrm{Droplets}} \right) dA_\textrm{r}~\qquad
    \label{eq_app:MEP}
\end{eqnarray}
where $T_\textrm{d}$ is the operating temperature ({\it i.~e.}, mean of near-surface ($T_{\rm{2m}}$) and skin temperature ($T_{\textrm{s}}$)),  $T_{\textrm{BL}}$ is the temperature at the boundary layer, $T_{\textrm{c}}$ is the working temperature at condensation, $T_{\textrm{p}}$ is the mean of $T_{\rm{c}}$ and $T_{\rm{s}}$, $\kappa_{\rm{s}}$ is the specific kinetic energy dissipation rate, $g$ is the gravitational acceleration constant and $h_{\rm{ct}}$ is the distance covered by droplets.
Contributions are integrated over the Earth's surface area $A$ or precipitation area $A_{\rm{r}}$. MEP associated to evaporation is directly derived from $E$ given by the MITgcm diagnostics and not from the latent heat. 

Note that material entropy production can also be estimated through an `indirect method' separating vertical and horizontal energy transport terms~\citep{10.1175/2011JAS3713.1,lembo_2019}. However, we have checked that in MITgcm such indirect method always overestimates vertical contributions. The low number of atmospheric level in MITgcm can be the source of this discrepancy with respect to the results obtained with the direct method.

\subsection{Lorenz Energy Cycle (LEC)}

Storage terms and conversion terms of LEC are computed in TheDiaTo using formulae given in Appendix~A of \citet{lembo_2019} or in \citet{ulbrich_global_1991}.
Generation and dissipation terms are computed as residuals of the conversion
terms at each reservoir.

\subsection{Horizontal and vertical interpolation}

TheDiaTo requires fields on a longitude-latitude grid. Thus we have interpolated the MITgcm output fields from the cubed-sphere grid with $32\times 32$ points per face (corresponding to an average  horizontal resolution of 2.8$^\circ$) to a longitude-latitude grid at 2$^\circ$.

MITgcm uses an Arakawa C-grid where scalar fields, like temperature and humidity, are saved at the center of grid cells, while vectorial fields are stored at the boundaries. This implies that, while all the scalar quantities and the horizontal components of wind speed  are calculated at the centers $k$, the vertical component  is computed at levels $k - 1/2$, with the lowest on the sea surface. Since TheDiaTo requires that all the fields are calculated at the same point, we have linearly interpolated the vertical component of the wind speed on the same levels $k$ as the other quantities.
For the upper level (75 hPa), we have used a null vertical component at TOA. 


\bibliographystyle{ametsoc2014}
\bibliography{attractors.bib}

\end{document}